\documentclass[a4paper,11pt]{article}
\pdfoutput=1 

\usepackage{jcappub} 

\usepackage[T1]{fontenc} 

\title{\boldmath Observing remnants by fermions' tunneling}

\author[a]{Deyou Chen,}
\author[b]{Houwen Wu,}
\author[b]{Haitang Yang}

\affiliation[a]{College of Physics and Electronic Information, China West Normal University, \\Nanchong 637009, China}
\affiliation[b]{Center for Theoretical Physics, College of Physical Science and Technology,
 \\Sichuan University, Chengdu 610064, China}

\emailAdd{dchen@cwnu.edu.cn}
\emailAdd{2013222020003@stu.scu.edu.cn}
\emailAdd{hyanga@scu.edu.cn}

\abstract{The standard Hawking formula predicts the complete evaporation of black holes. In this paper, we introduce effects of quantum gravity into fermions' tunneling from Reissner-Nordstrom and Kerr black holes. The quantum gravity effects slow down the increase of Hawking temperatures. This property naturally leads to a residue mass in black hole evaporation. The corrected temperatures are affected by the quantum numbers of emitted fermions. Meanwhile, the temperature of the Kerr black hole is a function of $\theta$ due to the rotation.}

\begin{document}
\maketitle
\flushbottom

\section{Introduction}
\label{Introduction}
Hawking radiation is interpreted as a quantum tunneling process at black holes' horizons. In the original research \cite{SWH}, the standard Hawking formula was derived. It implies that the complete evaporation of black holes. This result can be seen as a direct consequence of the Heisenberg uncertainty princplie.

The semi-classical tunneling method is an effective way to study the radiation \cite{KW}. Using this method, much fruit has been achieved \cite{PW,ZZ,JWC, KM,KSAE,LRC}. Taking into account the dynamics of spacetime, Parikh and Wilczek first researched the tunneling radiation of massless scalar particles in spherically symmetrical black holes \cite{PW}. The result shows that the tunneling rate is related to the change of Bekenstein-Hawking entropy. The corrected temperature is higher than the standard one, which implies that the varied spacetime accelerates the black holes' evaporation. Equations of motion of massless and massive particles have different features. Massless particles move along the null geodesics. The motion of massive particles obeys de Broglie wave and  is the phase velocity of outgoing particles. Thus using the relation between phase velocity and group velocity, this work was extended to the tunneling radiation of massive and charged scalar particles \cite{ZZ,JWC}. This result is full in consistence with that of Parikh and Wilczek. Subsequently, Kerner and Mann successfully extended this work to the radiation of fermions \cite{KM}. The standard Hawking temperature was recovered. In this work, one doesn't need the assumption that the particle move along the radial geodesics.

On the other hand, various theories of quantum gravity imply the existence of a minimal observable length \cite{PKT,ACV,KPP,LJG,GAC}. This length can be realized in the model of generalized uncertainty principle (GUP), which in turn is a consequence of the modified fundamental commutation relations. To study quantum effects of black holes, the traditional semi-classical methods quantize the emitted particle fields only and leave the gravitational background in a classical manner. While, the GUP model, introducing gravitational effects into quantum mechanics, is a different direction towards the quantum theory of gravity. The general form of the modified commutator, completely irrelevant to the  emitted particles, is a representation of the quantum property of gravity itself. Therefore, in some sense, the GUP model is a simple realization of the quantization of gravity, but not the emitted particles, though the  modifications are imposed upon the emitted particles in calculations.

Introducing  GUP into the black hole physics \cite{ACS,BG,BJM,SA,LW,MAJ,ZDM,NS}, many interest results have been discovered. In \cite{BG}, the remnant mass, corrections to the area law and heat capacity were obtained. A model for quantum black holes was introduced in \cite{BJM} and the authors showed that the Wheeler-Dewitt equation is similar to the equation of motion of a one-dimensional harmonic oscillator. Then the entropy and Hawking temperature of Schwarzschild black hole were addressed with GUP.  The black hole thermodynamics and the remnants were discussed in \cite{SA,LW,MAJ,ZDM}. In \cite{NS}, the authors modified the commutation relation between the radial coordinate and its conjugate momentum by the expression of GUP. They considered the existence of natural cutoffs as a minimal length, a minimal momentum and a maximal momentum. Then by combining Parikh-Wilczek semi-classical tunneling method and GUP, the radiation of massless
scalar particles in the Schwarzschild black hole was discussed. It turns out that the corrected Hawking temperature is dependent on the energy of emitted particles. However, all the above researches focus on scalar particles and to our knowledge, there is little discussion about fermions in literature.

In this paper, we investigate remnants by fermions' tunneling across the horizons of Reissner-Nordstrom and Kerr black holes. In the discussion, effects of quantum gravity are taken into account. Our calculation shows that the quantum gravity correction is related not only to the black hole's mass but also to the quantum numbers of emitted fermions. Moreover, the quantum gravity correction explicitly retards the temperature rising in the process of black hole evaporation. Therefore, at some point during the evaporation, the quantum correction balances the traditional temperature rising tendency. This leads to the existence of the remnants.

The organization of this paper is as follows. In section 2, from the modified fundamental commutation relation, we generalize Dirac equation in curved spacetime. In section 3, incorporating  GUP, we investigate the tunneling of charged fermions in the Reissner-Nordstrom black hole. The tunneling of uncharged fermions
in the Kerr black hole is discussed and the remnants are derived in section 4. Section 5 is devoted to our discussion and conclusion.

\section{Generalized Dirac equation}

An important model to realize the minimal observable length is the GUP

\begin{eqnarray}
\Delta x \Delta p \geq \frac{\hbar}{2}\left[1+ \beta \Delta
p^2\right], \label{eq2.1}
\end{eqnarray}

\noindent  where $\beta =\beta_0/M_p^2$. $M_p$ is the Planck mass.
$\beta_0 $ is a dimensionless parameter marking quantum gravity
effects. We set $c=G=k_B=1$ in this paper. We relaxing the upper
bound of $\beta_0$ from simple electroweak consideration $\beta_0
< 10^{5}$ \cite{LMS}. Kempf et. al. \cite{KMM} first made modifications
on the commutation relations $\left[x_i,p_j\right]= i \hbar
\delta_{ij}\left[1+ \beta p^2\right]$, where $x_i$ and $p_i$ are
position and momentum operators defined by

\begin{eqnarray}
x_i &=& x_{0i}, \nonumber\\
p_i &=& p_{0i} (1 + \beta p^2), \label{eq2.2}
\end{eqnarray}

\noindent respectively. $x_{0i}$ and $p_{0j}$ satisfy the
canonical commutation relations $\left[x_{0i},p_{0j}\right]= i
\hbar \delta_{ij}$. Then one gets

\begin{eqnarray}
p^2 &=& p_i p^i = -\hbar^2 \left[ {1 - \beta \hbar^2 \left(
{\partial _j \partial ^j} \right)} \right]\partial _i \cdot \left[
{1 - \beta \hbar^2 \left( {\partial ^j\partial _j } \right)}
\right]\partial ^i\nonumber \\
&\simeq & - \hbar ^2\left[ {\partial _i \partial ^i - 2\beta \hbar
^2 \left( {\partial ^j\partial _j } \right)\left( {\partial
^i\partial _i } \right)} \right]. \label{eq2.3}
\end{eqnarray}

\noindent In the last step, only leading order term of $\beta $ is
kept. Following  \cite{WG}, to realize quantum gravity effects,
the definition of generalized frequency is found to be

\begin{eqnarray}
\tilde \omega = E( 1 - \beta E^2), \label{eq2.4}
\end{eqnarray}

\noindent with the definition of energy operator $ E = i \hbar
\partial _0 $. Considering the energy mass shell condition $ p^2 +
m^2 = E^2 $, we get the expression of energy
\cite{NS,WG,NK,HBHRSS}

\begin{eqnarray}
\tilde E = E[ 1 - \beta (p^2 + m^2)]. \label{eq2.5}
\end{eqnarray}

\noindent  For massless particles, $m = 0$, the tunneling in
Schwarzschild spacetime was studied and the corrected black hole's
temperature was given  in \cite{NS}. In this paper, we investigate
the radiation of spin-1/2 fermions in curved spacetime where
effects of quantum gravity are taken into account. Generalized
Dirac equation based on GUP in  flat spacetime has been gotten in
\cite{NK}. In curved spacetime, Dirac equation with an
electromagnetic field is

\begin{eqnarray}
i\gamma^{\mu}\left(\partial_{\mu}+\Omega_{\mu}+\frac{i}{\hbar}eA_{\mu}\right)\psi+\frac{m}{\hbar}\psi=0,
\label{eq2.6}
\end{eqnarray}

\noindent where $\Omega _\mu \equiv\frac{i}{2}\omega _\mu\, ^{a b}
\Sigma_{ab}$, $\omega _\mu\, ^{ab}$ is the spin connection defined
by the ordinary connection and the tetrad $e^\lambda\,_b$

\begin{equation}
\omega_\mu\,^a\,_b=e_\nu\,^a e^\lambda\,_b \Gamma^\nu_{\mu\lambda}
-e^\lambda\,_b\partial_\mu e_\lambda\,^a.
\end{equation}

\noindent The Greek indices are raised and lowered by the curved
metric $g_{\mu\nu}$. The Latin indices are governed by the flat
metric $\eta_{ab}$. To construct the tetrad, one uses the
following definitions,

\begin{equation}
g_{\mu\nu}= e_\mu\,^a e_\nu\,^b \eta_{ab},\hspace{5mm} \eta_{ab}=
g_{\mu\nu} e^\mu\,_a e^\nu\,_b, \hspace{5mm} e^\mu\,_a e_\nu\,^a=
\delta^\mu_\nu, \hspace{5mm} e^\mu\,_a e_\mu\,^b = \delta_a^b.
\label{eq6-1}
\end{equation}

\noindent In the definition of $\Omega_\mu$, $\Sigma_{ab}$'s are
the Lorentz spinor generators defined by

\begin{equation}
\Sigma_{ab}= \frac{i}{4}\left[ {\gamma ^a ,\gamma^b} \right],
\hspace{5mm} \{\gamma ^a ,\gamma^b\}= 2\eta^{ab}. \label{eq6-2}
\end{equation}

\noindent Therefore, it is readily to construct the $\gamma^\mu$'s
in curved spacetime as

\begin{equation}
\gamma^\mu = e^\mu\,_a \gamma^a, \hspace{7mm} \left\{ {\gamma ^\mu
,\gamma ^\nu } \right\} = 2g^{\mu \nu }. \label{eq6-3}
\end{equation}

\noindent Equation (\ref{eq2.6}) can be rewritten as

\begin{eqnarray}
-i\gamma^{0}\partial_{0}\psi=\left(i\gamma^{i}\partial_{i}+i\gamma^{\mu}\Omega_{\mu}+i\gamma^{\mu}
\frac{i}{\hbar}eA_{\mu}+\frac{m}{\hbar}\right)\psi. \label{eq2.7}
\end{eqnarray}

\noindent Inserting equations (\ref{eq2.3}) and (\ref{eq2.5}) into
equation (\ref{eq2.7}) and neglecting higher orders of $\beta$
yield the generalized Dirac equation in curved spacetime

\begin{eqnarray}
-i\gamma^{0}\partial_{0}\psi=\left(i\gamma^{i}\partial_{i}+i\gamma^{\mu}\Omega_{\mu}+i\gamma^{\mu}
\frac{i}{\hbar}eA_{\mu}+\frac{m}{\hbar}\right)\left(1+\beta\hbar^{2}\partial_{j}\partial^{j}-\beta
m^{2}\right)\psi, \label{eq2.8}
\end{eqnarray}

\noindent which can be rewritten as

\begin{eqnarray}
\left[i\gamma^{0}\partial_{0}+i\gamma^{i}\partial_{i}\left(1-\beta
m^{2}\right)+i\gamma^{i}\beta\hbar^{2}\left(\partial_{j}\partial^{j}\right)\partial_{i}+\frac{m}{\hbar}
\left(1+\beta\hbar^{2}\partial_{j}\partial^{j}-\beta m^{2}\right)\right.\nonumber \\
\left.+i\gamma^{\mu}\frac{i}{\hbar}eA_{\mu}\left(1+\beta\hbar^{2}\partial_{j}\partial^{j}-\beta
m^{2}\right)+i\gamma^{\mu}\Omega_{\mu}\left(1+\beta\hbar^{2}\partial_{j}\partial^{j}-\beta
m^{2}\right)\right]\psi= 0. \label{eq2.9}
\end{eqnarray}

\noindent This is the equation of motion of charged fermions. When
$e= 0$, it describes the motion of uncharged fermions.  In the
following sections, equation (\ref{eq2.9}) is adopted to discuss
the tunneling radiation of fermions in the Reissner-Nordstrom and
the Kerr spacetimes.

\section{Fermion's tunneling in the Reissner-Nordstrom spacetime}

The Reissner-Norstrom black hole describes a spherrically
symmetric static spacetime with charge $Q$. The metric is given by

\begin{eqnarray}
ds^{2}=-f\left(r\right)dt^{2}+g\left(r\right)^{-1}dr^{2}+r^{2}\left(d\theta^{2}+\sin^{2}\theta
d\phi^{2}\right), \label{eq3.1}
\end{eqnarray}

\noindent with the electromagnetic potential

\begin{eqnarray}
A_{\mu}=\left(A_{t},0,0,0\right)= \left(\frac{Q}{r},0,0,0\right),
\label{eq3.2}
\end{eqnarray}

\noindent where

\begin{eqnarray}
f\left(r\right)=g\left(r\right)=1-\frac{2M}{r}+\frac{Q^{2}}{r^{2}}=
\frac{(r-r_+)(r-r_-)}{r^2}. \label{eq3.3}
\end{eqnarray}

\noindent $r_\pm = M \pm \sqrt{M^2-Q^2}$  are locations of the
outer horizon and the inner horizon, respectively. For a spin-1/2
particle, there are two states corresponding to spin up and spin
down. In this paper, without losing generality, we only consider
the state with spin up. The calculation of the spin down state is
parallel. Therefore, we suppose  the wave function of the emitted
fermions is

\begin{eqnarray}
\Psi=\left(\begin{array}{c}
A\\
0\\
B\\
0
\end{array}\right)\exp\left(\frac{i}{\hbar}I\left(t,r,\theta,\phi\right)\right),
\label{eq3.4}
\end{eqnarray}

\noindent where $I$ is the action and $A$, $B$ are functions of
$t, r , \theta , \phi$. There are many choices to construct the
$\gamma^\mu$ matrices. We first explore the tetrad.  For the
metric (\ref{eq3.1}), one can easily construct

\begin{eqnarray}
e_\mu\,^a = \rm{diag}\left(\sqrt f, 1/\sqrt g, r, r\sin\theta
\right).
\end{eqnarray}

\noindent Then the $\gamma^\mu$ matrices are

\begin{eqnarray}
\gamma^{t}=\frac{1}{\sqrt{f\left(r\right)}}\left(\begin{array}{cc}
i & 0\\
0 & -i
\end{array}\right), &  & \gamma^{\theta}=\sqrt{g^{\theta\theta}}\left(\begin{array}{cc}
0 & \sigma^{1}\\
\sigma^{1} & 0
\end{array}\right),\nonumber \\
\gamma^{r}=\sqrt{g\left(r\right)}\left(\begin{array}{cc}
0 & \sigma^{3}\\
\sigma^{3} & 0
\end{array}\right), &  & \gamma^{\phi}=\sqrt{g^{\phi\phi}}\left(\begin{array}{cc}
0 & \sigma^{2}\\
\sigma^{2} & 0
\end{array}\right).
\label{eq3.5}
\end{eqnarray}

\noindent In the equation above, $\sigma ^i$'s are the Pauli
matrices, $\sqrt{g^{\theta\theta}} =\frac{1}{r}$ and
$\sqrt{g^{\phi\phi}} =\frac{1}{r\sin\theta}$. Inserting the wave
function and gamma matrices into the generalized Dirac equation
(\ref{eq2.9}),  applying the WKB approximation, and keep only the
leading order of $\hbar$, we get the the equations of motion

\begin{eqnarray}
-iA\frac{1}{\sqrt{f}}\partial_{t}I-B\left(1-\beta
m^{2}\right)\sqrt{g}\partial_{r}I-Am\beta\left[g^{rr}\left(\partial_{r}I\right)^{2}+g^{\theta\theta}\left(\partial_{\theta}I\right)^{2}+
g^{\phi\phi}\left(\partial_{\phi}I\right)^{2}\right]\nonumber\\
+B\beta\sqrt{g}\partial_{r}I
\left[g^{rr}\left(\partial_{r}I\right)^{2}+g^{\theta\theta}\left(\partial_{\theta}I\right)^{2}
+g^{\phi\phi}\left(\partial_{\phi}I\right)^{2}\right]+Am\left(1-\beta m^{2}\right)\nonumber\\
-iA\frac{eA_t}{\sqrt f}\left[1-\beta m^{2}
-\beta\left(g^{rr}\left(\partial_{r}I\right)^{2}+g^{\theta\theta}\left(\partial_{\theta}I\right)^{2}+
g^{\phi\phi}\left(\partial_{\phi}I\right)^{2}\right)\right] = 0,
\label{eq3.6}
\end{eqnarray}

\begin{eqnarray}
iB\frac{1}{\sqrt{f}}\partial_{t}I-A\left(1-\beta
m^{2}\right)\sqrt{g}\partial_{r}I-Bm\beta\left[g^{rr}\left(\partial_{r}I\right)^{2}+g^{\theta\theta}\left(\partial_{\theta}I\right)^{2}+
g^{\phi\phi}\left(\partial_{\phi}I\right)^{2}\right]\nonumber\\
+A\beta\sqrt{g}\partial_{r}I
\left[g^{rr}\left(\partial_{r}I\right)^{2}+g^{\theta\theta}\left(\partial_{\theta}I\right)^{2}
+g^{\phi\phi}\left(\partial_{\phi}I\right)^{2}\right]+Bm\left(1-\beta m^{2}\right)\nonumber\\
+iB\frac{eA_t}{\sqrt f}\left[1-\beta m^{2}
-\beta\left(g^{rr}\left(\partial_{r}I\right)^{2}+g^{\theta\theta}\left(\partial_{\theta}I\right)^{2}+
g^{\phi\phi}\left(\partial_{\phi}I\right)^{2}\right)\right] = 0,
\label{eq3.7}
\end{eqnarray}

\begin{eqnarray}
A\left\{-(1-\beta m^2) \sqrt {g^{\theta \theta}}\partial _
{\theta} I + \beta \sqrt {g^{\theta \theta}}\partial _
{\theta}I\left[g^{rr}(\partial _r I)^2 + g^{\theta
\theta}(\partial _ {\theta}I)^2
+ g^{\phi \phi}(\partial _ {\phi}I)^2 \right]\right.\nonumber\\
\left.-i(1-\beta m^2) \sqrt {g^{\phi \phi}}\partial _ {\phi}I+
i\beta \sqrt {g^{\phi \phi}}\partial _ {\phi}I\left[g^{rr}
(\partial _r I)^2 + g^{\theta \theta}(\partial _ {\theta}I)^2 +
g^{\phi \phi}(\partial _ {\phi}I)^2\right]\right\} = 0,
\label{eq3.8}
\end{eqnarray}

\begin{eqnarray}
B\left\{-(1-\beta m^2) \sqrt {g^{\theta \theta}}\partial _
{\theta} I + \beta \sqrt {g^{\theta \theta}}\partial _
{\theta}I\left[g^{rr}(\partial _r I)^2 + g^{\theta
\theta}(\partial _ {\theta}I)^2
+ g^{\phi \phi}(\partial _ {\phi}I)^2 \right]\right.\nonumber\\
\left.-i(1-\beta m^2) \sqrt {g^{\phi \phi}}\partial _ {\phi}I+
i\beta \sqrt {g^{\phi \phi}}\partial _ {\phi}I\left[g^{rr}
(\partial _r I)^2 + g^{\theta \theta}(\partial _ {\theta}I)^2 +
g^{\phi \phi}(\partial _ {\phi}I)^2\right]\right\} = 0.
\label{eq3.9}
\end{eqnarray}

\noindent  It is difficult to solve the action from the above
equations. Considering the property of the Reissner-Nordstrom
spacetime and the question we are addressing, following the
standard process, we carry out separation of variables

\begin{eqnarray}
I = -\omega t + W(r) + \Theta (\theta , \phi), \label{eq3.10}
\end{eqnarray}

\noindent where $\omega$ is the energy of emitted fermions. We
first observe the last two equations in
(\ref{eq3.6})-(\ref{eq3.9}). Inserting equation (\ref{eq3.10})
into equations (\ref{eq3.8}) and (\ref{eq3.9}), cancelling
respectively  $A$ and $B$, we find that they are the same equation
and can be written as

\begin{eqnarray}
\left(\sqrt {g^{\theta \theta}}\partial _ {\theta} \Theta +i\sqrt
{g^{\phi \phi}}\partial _ {\phi}\Theta\right) \left[\beta
g^{rr}(\partial _r W)^2 + \beta g^{\theta \theta}(\partial _
{\theta}\Theta)^2 +\beta g^{\phi \phi}(\partial _ {\phi}\Theta)^2
+\beta m^2 -1\right] =0, \label{eq3.11}
\end{eqnarray}

\noindent which implies

\begin{eqnarray}
\sqrt {g^{\theta \theta}}\partial _ {\theta} \Theta +i\sqrt
{g^{\phi \phi}}\partial _ {\phi}\Theta =0, \label{eq3.12}
\end{eqnarray}

\noindent since the terms in the square bracket can not be
balanced to vanish. The solution of $\Theta $ is a complex
function (other than the trivial one $\Theta = constant$) and has
contribution to the action. However, it has no contribution to the
tunneling rate.
%
%
%
%
Now we consider equations (\ref{eq3.6}) and (\ref{eq3.7}), from
which the radial action is derived and the temperature of black
hole is determined. Substituting equation (\ref{eq3.10}) into
equations (\ref{eq3.6}), (\ref{eq3.7}) and canceling $A$ and $B$
yield

\begin{eqnarray}
A_6\left(\partial_{r}W\right)^{6}+A_4\left(\partial_{r}W\right)^{4}+A_2\left(\partial_{r}W\right)^{2}+A_0=0,
\label{eq3.14}
\end{eqnarray}

\noindent where

\begin{eqnarray}
A_6 &=& \beta^{2}g^{3}f,\nonumber\\
A_4 &=& \beta g^{2}f\left(m^{2}\beta-2\right)-\beta ^2 g^2 e^2A_t^2,\nonumber\\
A_2 &=& gf\left(1-\beta m^{2}\right)\left(1+\beta m^{2}\right)+2\beta g eA_t [-\omega + eA_t(1-\beta m^2)],\nonumber\\
A_0 &=& -m^{2}f\left(1-\beta m^{2}\right)^{2}-\left[\omega
-eA_t\left(1-\beta m^{2}\right)\right]^2. \label{eq3.15}
\end{eqnarray}

\noindent Neglecting higher order terms of $\beta$ and solving
equation (\ref{eq3.14}) at the event horizon yield the solution of
the radial action. The particle's tunneling rate is determined by
the imaginary part of the action,

\begin{eqnarray}
Im W_\pm (r) & = &Im \int dr\frac{1}{\sqrt{gf}}\sqrt{m^{2}f+\left[\omega
- e A_{t} (1-\beta m^2)\right]^2}
\left(1+\beta m^{2}+\beta\frac{\tilde \omega_0^{2}}{f}- \frac{\beta eA_t \tilde\omega_0}{f}\right) \nonumber \\
& = & \pm \pi\frac{r_+^2}{r_+ - r_-} (\omega - eA_{t+})
\times\left(1+ \beta\Xi\right) ,
 \label{eq3.16}
\end{eqnarray}

\noindent where $+(-)$ are solutions of outgoing (ingoing) waves,
$\tilde\omega_0 = \omega - eA_{t}$ and $A_{t+} = \frac{Q}{r_+}$ is
the electromagnetic potential at the event horizon. $\Xi$ is
given by

\begin{eqnarray}
\Xi & = &\frac{3m^2}{2}+\frac{em^2A_{t+}}{\omega - eA_{t+}} + 2\frac{4e\omega Qr_+r_- + \omega^2 r_+^3 -2e^2Q^2(r_+ + r_-) -e\omega Q r_+^2 - 2\omega^2r_+^2r_-}{\left(r_+-r_-\right)^2r_+}.
\label{eq3.17}
\end{eqnarray}

\noindent Using $r_{\pm} = M \pm \sqrt{M^2-Q^2}$, The right hand side of eqn. (\ref{eq3.17}) is reduced into $\frac{3m^2}{2}+\frac{em^2A_{t+}}{\omega - eA_{t+}}+ 2\frac{2eQ^2\left(\omega Q -eM \right) + \omega r_+ \left(\omega - eA_{t+}\right)\left(r_+^2-Q^2\right)}{\left(r_+-r_-\right)^2r_+}$. It is easily found that $\Xi >0$. Thus the tunneling rate
of fermions at the event horizon is

\begin{eqnarray}
\Gamma & = & \frac{P_{\rm (emission)}}{P_{\rm(absorption)}} =
\frac{\exp\left(
 -2\,\mathrm{Im} W_+ - 2\,\mathrm{Im} {\Theta}\right)}{\exp\left( -2\,
 \mathrm{Im} W_- -2\,\mathrm{Im} {\Theta}\right)}\nonumber\\
& = & \exp\left[ -4\pi\frac{r_+^2}{r_+ - r_-} (\omega - eA_{t+})
\times\left(1 + \beta\Xi\right)\right]. \label{eq3.18}
\end{eqnarray}

\noindent This is the Boltzmann factor for an object with the
effective temperature

\begin{eqnarray}
T = \frac{r_+ - r_-}{4\pi r_+^2\left(1+
\beta\Xi\right)}= T_0\left(1 -\beta\Xi\right) ,
\label{eq3.19}
\end{eqnarray}

\noindent where $T_0 =  \frac{r_+ - r_-}{4\pi r_+^2}$ is the
original Hawking temperature of the Reissner-Nordstrom black hole.
Therefore, the corrected temperature relies on the quantum numbers
(charge, mass, energy) of the emitted fermions. Moreover, the
quantum effects explicitly decelerates the temperature increase
during the evaporation. Thus, it is conceivable that the two
tendencies will be cancelled at some point in the radiation and
remnants are left.

\section{Fermion's tunneling in the Kerr spacetime}

In this section, we investigate the fermion's tunneling at the
event horizon of the Kerr black hole where GUP is taken into
account. Here for simplicity, we suppose the emitted fermions are
uncharged, therefore we set the electromagnetic field charge
vanishing in equation (\ref{eq2.9}). The Kerr metric is given by

\begin{eqnarray}
ds^{2} &=& - \left(1-\frac{2Mr}{\rho^{2}}\right) dt^2+\frac{\rho^{2}}{\Delta}dr^{2}
+\left[ (r^2+a^2) + \frac{2Mra^2\sin^2 {\theta}}{\rho^{2}}\right]\sin^2 {\theta}d\varphi^2 \nonumber\\
&& +\rho^2 d\theta^2-\frac{4Mra\sin^2
{\theta}}{\rho^{2}}dtd\varphi, \label{eq4.1}
\end{eqnarray}

\noindent where

\begin{eqnarray}
\rho^{2} &=& r^2 + a^2\cos^2{\theta},\nonumber\\
\Delta &=& r^2 - 2Mr +a^2 = (r- r_+)(r-r_-). \label{eq4.2}
\end{eqnarray}

\noindent $r_\pm = M \pm \sqrt{M^2-a^2}$  are locations of the
outer and inner horizons. $M$ is the black hole's mass and $a$ is
the angular momentum per unit mass. To calculate the fermion's
tunneling, one can directly construct the $\gamma^\mu$ matrices
from the metric (\ref{eq4.1}). One of such constructions can be
found in \cite{KM}. For convenience, we perform the dragging
coordinate transformation $\phi = \varphi - \Omega t$ on the
metric (\ref{eq4.1}), where

\begin{eqnarray}
\Omega = \frac{(r^2+a^2-\Delta)a}{\left(r^2+a^2\right)^2 - \Delta
a^2\sin^2{\theta}}, \label{eq4.3}
\end{eqnarray}

\noindent is the black hole's angular velocity. Then we get

\begin{eqnarray}
ds^{2}&=& -\frac{\triangle\rho^{2}}{\left(r^{2}+a^{2}\right)^{2}-\triangle a^{2}
\sin^{2}\theta}dt^{2}+\frac{\rho^{2}}{\triangle}dr^{2}+\rho^{2}d\theta^{2}\nonumber\\
&& +\frac{\left(r^{2}+a^{2}\right)^{2}-\triangle
a^{2}\sin^{2}\theta}{\rho^{2}}\sin^{2}\theta d\phi^{2}\nonumber\\
 & \equiv& -F(r) dt^2 + \frac{1}{G(r)}dr^2 + K^2(r) d {\theta}^2 + H^2(r)d
 {\phi}^2.
\label{eq4.4}
\end{eqnarray}

\noindent From the comparability between the metric (\ref{eq4.4})
and the metric (\ref{eq3.1}), a tetrad for the metric
(\ref{eq4.4}) can be figured out $e_\mu\,^a = \rm{diag}\left(\sqrt
F, 1/\sqrt G, K, H\right)$. Then the gamma matrices are given by

\begin{eqnarray}
\gamma^{t}=\frac{1}{\sqrt{F\left(r\right)}}\left(\begin{array}{cc}
i & 0\\
0 & -i
\end{array}\right), &  & \gamma^{\theta}=\frac{1}{K\left(r\right)}\left(\begin{array}{cc}
0 & \sigma^{1}\\
\sigma^{1} & 0
\end{array}\right),\nonumber \\
\gamma^{r}=\sqrt{G\left(r\right)}\left(\begin{array}{cc}
0 & \sigma^{3}\\
\sigma^{3} & 0
\end{array}\right), &  & \gamma^{\phi}=\frac{1}{H\left(r\right)}\left(\begin{array}{cc}
0 & \sigma^{2}\\
\sigma^{2} & 0
\end{array}\right).
\label{eq4.5}
\end{eqnarray}



\noindent We again only need to  consider the spin up states.
Therefore the wave function  (\ref{eq3.4}) also applies to our
current calculation. Similar to what we did in the last section,
after inserting the wave function and the gamma matrices into the
generalized uncharged Dirac equation, we get four equations

\begin{eqnarray}
-iA\frac{1}{\sqrt{F}}\partial_{t}I-B\left(1-\beta
m^{2}\right)\sqrt{G}\partial_{r}I-Am\beta\left[g^{rr}\left(\partial_{r}I\right)^{2}+g^{\theta\theta}\left(\partial_{\theta}I\right)^{2}+
g^{\phi\phi}\left(\partial_{\phi}I\right)^{2}\right]\nonumber\\
+B\beta\sqrt{G}\partial_{r}I\left[g^{rr}\left(\partial_{r}I\right)^{2}+g^{\theta\theta}\left(\partial_{\theta}I\right)^{2}+
g^{\phi\phi}\left(\partial_{\phi}I\right)^{2}\right]+Am\left(1-\beta
m^{2}\right) = 0, \label{eq4.7}
\end{eqnarray}

\begin{eqnarray}
iB\frac{1}{\sqrt{F}}\partial_{t}I-A\left(1-\beta
m^{2}\right)\sqrt{G}\partial_{r}I-Bm\beta\left[g^{rr}\left(\partial_{r}I\right)^{2}+g^{\theta\theta}\left(\partial_{\theta}I\right)^{2}+
g^{\phi\phi}\left(\partial_{\phi}I\right)^{2}\right]\nonumber\\
+A\beta\sqrt{G}\partial_{r}I\left[g^{rr}\left(\partial_{r}I\right)^{2}+g^{\theta\theta}\left(\partial_{\theta}I\right)^{2}+
g^{\phi\phi}\left(\partial_{\phi}I\right)^{2}\right]+Bm\left(1-\beta
m^{2}\right) = 0, \label{eq4.8}
\end{eqnarray}

\begin{eqnarray}
A\left\{-\left(1-\beta
m^{2}\right)\frac{1}{K}\partial_{\theta}I+\beta\frac{1}{K}\partial
_{\theta}I\left[g^{rr}\left(\partial_{r}I\right)^{2}+g^{\theta\theta}\left(\partial_{\theta}I\right)^{2}+
g^{\phi\phi}\left(\partial_{\phi}I\right)^{2}\right]\right.\nonumber\\
\left.-i\left(1-\beta
m^{2}\right)\frac{1}{H}\partial_{\phi}I+i\beta\frac{1}{H}\partial
_{\phi}I\left[g^{rr}\left(\partial_{r}I\right)^{2}+g^{\theta\theta}\left(\partial_{\theta}I\right)^{2}+
g^{\phi\phi}\left(\partial_{\phi}I\right)^{2}\right]\right\} = 0,
\label{eq4.9}
\end{eqnarray}

\begin{eqnarray}
B\left\{-\left(1-\beta
m^{2}\right)\frac{1}{K}\partial_{\theta}I+\beta\frac{1}{K}\partial
_{\theta}I\left[g^{rr}\left(\partial_{r}I\right)^{2}+g^{\theta\theta}\left(\partial_{\theta}I\right)^{2}+
g^{\phi\phi}\left(\partial_{\phi}I\right)^{2}\right]\right.\nonumber\\
\left.-i\left(1-\beta
m^{2}\right)\frac{1}{H}\partial_{\phi}I+i\beta\frac{1}{H}\partial
_{\phi}I\left[g^{rr}\left(\partial_{r}I\right)^{2}+g^{\theta\theta}\left(\partial_{\theta}I\right)^{2}+
g^{\phi\phi}\left(\partial_{\phi}I\right)^{2}\right]\right\} = 0.
\label{eq4.10}
\end{eqnarray}

\noindent We then carry out separation of variables as

\begin{eqnarray}
I=-\left(\omega -j \Omega \right)t+ W\left(r, \theta \right) +j
\phi, \label{eq4.11}
\end{eqnarray}

\noindent where $\omega$ and $j$ are the energy and angular
momentum of emitted fermions, respectively. We first observe
equations (\ref{eq4.9}) and (\ref{eq4.10}). It turns out they are
identical and can be   rewritten as

\begin{eqnarray}
\left(\frac{1}{K}\partial_{\theta}I +
i\frac{1}{H}\partial_{\phi}I\right)\left[\beta
g^{rr}\left(\partial_{r}I\right)^{2}+\beta
g^{\theta\theta}\left(\partial_{\theta}I\right)^{2}+\beta
g^{\phi\phi}\left(\partial_{\phi}I\right)^{2}+\beta
m^{2}-1\right]=0, \label{eq4.12}
\end{eqnarray}

\noindent which reduces to

\begin{eqnarray}
\frac{1}{K}\partial_{\theta}I + i\frac{1}{H}\partial_{\phi}I=0,
\label{eq4.13}
\end{eqnarray}

\noindent and implies $
g^{\theta\theta}\left(\partial_{\theta}I\right)^{2}+g^{\phi\phi}\left(\partial_{\phi}I\right)^{2}
=0 $. In previous work, $ W(r,\theta)$ could be separated
furthermore as $W(r)\Theta(\theta)$. Here we still take the form
of $ W(r,\theta)$, fixing $\theta$ at a certain value $\theta_0$,
then solve $ W(r,\theta_0)$ at the event horizon \cite{KM2} from
equations (\ref{eq4.7}) and (\ref{eq4.8}). Substitute equation
(\ref{eq4.11}) into equations (\ref{eq4.7}) and (\ref{eq4.8}), we
get

\begin{eqnarray}
B_6\left(\partial_{r}W\right)^{6}+B_4\left(\partial_{r}W\right)^{4}+B_2\left(\partial_{r}W\right)^{2}+B_0=0,
\label{eq4.14}
\end{eqnarray}

\noindent where

\begin{eqnarray}
B_6 & = & \beta^{2}G^{3}F,\nonumber \\
B_4 & = & \beta G^{2}F\left(m^{2}\beta-2\right),\nonumber \\
B_2 & = & GF\left[\left(1-\beta m^{2}\right)^{2}+2\beta m^{2}\left(1-m^{2}\beta\right)\right],\nonumber \\
B_0 & = & -m^{2}\left(1-\beta
m^{2}\right)^{2}F-\left(\omega-j\Omega\right)^{2}. \label{eq4.15}
\end{eqnarray}

\noindent Neglecting higher order terms of $\beta$, we solve
equation (\ref{eq4.14}) at the event horizon and get

\begin{eqnarray}
W_{\pm} & = & \pm\int dr\sqrt{\frac{\left(\omega-j\Omega\right)^{2}+m^{2}F}{FG}}\left[1+\beta
\left(m^{2}+\frac{\left(\omega-j\Omega\right)^{2}}{F}\right)\right]\nonumber\\
 & = & \pm i\pi \left(\omega -j \Omega _+\right)\frac{r_+^2 +a^2}{r_+ - r_-}
 \left( 1+  \beta \Pi\right) + \hbox{(real part)},
\label{eq4.16}
\end{eqnarray}

\noindent where $+(-)$ are solutions of outgoing (ingoing) waves,
$\Omega_+ = \frac{a}{r_+^2 +a^2}$ is the angular velocity at the
event horizon, and the (real part) does not contribute to the
tunneling rate. The value of $\Pi$ is given by

\begin{eqnarray}
\Pi &=& \frac{3m^2}{2}- \frac{3\omega_0}{(r_+ - r_-)\rho_+^2}\left[j(r_+ + r_-)a - j\Omega_+\left(4(r_+^2+a^2)r_+ -(r_+ - r_-)a^2\sin^2{\theta_0}\right)\right]\nonumber\\
&& + \frac{\omega_0^2}{(r_+ - r_-)\rho_+^2}\left[12(r_+^2 + a^2)r_+ -3(r_+ - r_-)a^2\sin^2{\theta_0}\right.\nonumber\\
&& \left.- \frac{2(r_+^2 +a^2)^2}{r_+ - r_-}- \frac{2(r_+^2
+a^2)^2r_+}{\rho_+^2} \right], \label{eq4.17}
\end{eqnarray}

\noindent with $\omega_0 = \omega - j \Omega_+$, $\rho_+^2 = r_+^2
+ a^2 \cos^2{\theta_0}$. Using $r_{\pm} = M \pm \sqrt{M^2-a^2}$, it is again not hard to show that $\Pi$
is positive. Thus the tunneling rate of
uncharged fermions at the event horizon of Kerr black hole is

\begin{eqnarray}
\Gamma & = & \frac{P_{\rm(emission)}}{P_{\rm (absorption)}} = \frac{\exp
\left(-2\, \rm{Im}W_+ \right)}{\exp\left(-2\, \rm{Im} W_- \right)}\nonumber\\
& = & \exp\left[-4\pi\left(\omega -j \Omega _+\right)\frac{r_+^2
+a^2}{r_+ - r_-}\left( 1+ \beta \Pi\right)\right].
\label{eq4.18}
\end{eqnarray}

\noindent This is the expression of Boltzmann factor with a
temperature

\begin{eqnarray}
T =\frac{r_+ - r_-}{4\pi(r_+^2 +a^2)}\frac{1}{\left( 1+
\beta \Pi \right)} = T_0(1-  \beta \Pi), \label{eq4.19}
\end{eqnarray}

\noindent where $T_0 = \frac{r_+ - r_-}{4\pi(r_+^2 +a^2)} $ is the original
Hawking temperature of the Kerr black hole. Similar to the results
of Reissner-Nordstrom, the corrected temperature  is lower than
the original Hawking temperature and is related not only to the
black hole's mass and angular momentum, but also to the quantum
numbers (mass, angular momentum, energy) of emitted fermions. The
quantum effects also stop the temperature increase at a balance
point during the evaporation and leave remnants of the black hole.

When $\beta=0$, the the original Hawking temperatures of
Reissner-Nordstrom and Kerr black holes are recovered by Eqs.
(\ref{eq3.19}) and (\ref{eq4.19}), respectively. When $Q = 0$ and
$a=0$, the Reissner-Nordstrom metric and the Kerr metric are
reduced to the Schwarzschild metric. Then the corrected Hawking
temperature

\begin{eqnarray}
T = \frac{1}{8\pi M} \left[1-\frac{1}{2}\beta
\left(3m^{2}+4\omega^{2}\right)\right] \label{eq5.2}
\end{eqnarray}

\noindent is that of Schwarzschild black hole. To
estimate the residue mass, it is enough to consider massless
particles. To avoid the temperature $T$ becoming negative, the value of $\omega$ should  satisfy $\omega < \frac{M_p}{\sqrt{\beta_0}}$, where a factor of $2$ is omitted since we are only concerned with the order of magnitude. The temperature stops increasing when

\begin{equation}
(M-dM)(1+\beta \omega^2)\simeq M.
\end{equation}

\noindent Then using the condition $dM = \omega$ and $\beta =\beta_0/M_p^2$
where $M_p$ is the Planck mass and $\beta_0 <10^{5}$ \cite{LMS}
is a dimensionless parameter marking quantum gravity effects, we get

\begin{equation}
M_{\hbox{Res}} \simeq \frac{M_p^2}{\beta_0 \omega} \gtrsim
\frac{M_p}{\sqrt{\beta_0}}, \hspace{7mm} T_{\hbox{Res}} \lesssim
\frac{\sqrt{\beta_0}}{8\pi M_p}.
\label{final results}
\end{equation}

\noindent This result is consistent with those obtained in \cite{ACS,BG,CA}. Compared with previous results, our calculation explicitly shows how the residue mass of black holes arises due to quantum gravity
effects. It is known that the WKB-type of approximation is basically the same as working with a 1+1-dimensional spacetime. As a consequence, all large non-extremal black holes look basically the same (like Rindler space). It is also true for the Reissner-Nordstrom and the Kerr black holes. Our investigation in this paper shows this consequence \cite{IUW}.

Refer to (\ref{final results}), there is an upper bound of the temperature for the remnants. This temperature may be far above the Planck temperature since $\beta_0< 10^5$. It is not clear if the concept of temperature still holds beyond the Planck temperature. Since we are taking an effective model in this work, we expect a full quantum theory of gravity can answer this question.

\section{Discussion and conclusion}
In this paper, incorporating effects of quantum gravity, we derived the generalized Dirac equation in curved spacetime based on the modified fundamental commutation relations \cite{KMM}.  We investigated the fermions' tunneling in the Reissner-Nordstrom and Kerr black holes. In both spacetime configurations, we showed that the corrected Hawking temperature is not only determined by the properties of the black holes, but also dependent on the quantum numbers (charge, angular momentum, mass, energy) of the emitted particles. Our calculation implies that the temperature increasing during the evaporation is slowed down by the quantum effects. At some point, these two tendencies will be balanced and lead to remnants of the black holes. The remnants was derived as $M_{\hbox{Res}}\gtrsim\frac{M_p}{\sqrt{\beta_0}}$ by the emission of the massless particles.

\acknowledgments
We are very grateful for S.Q. Wu and  P. Wang for their useful discussions. This work is supported in part by the NSFC (Grant Nos. 11205125, 11175039, 11375121) and SiChuan Province Science Foundation for Youths (Grant No. 2012JQ0039).


\end{document}